\documentclass{article}

\usepackage{authblk}
\usepackage[english]{babel}
\usepackage[utf8x]{inputenc}
\usepackage{amsmath}
\usepackage{graphicx, subcaption}
\usepackage[colorinlistoftodos]{todonotes}
\usepackage{float}
\usepackage{hyperref}
\usepackage{cite}
\usepackage{tikz}
\usepackage{xfrac}
\usepackage{mathrsfs}
\usepackage{amsfonts}
\usepackage{amssymb}
\providecommand{\U}[1]{\protect\rule{.1in}{.1in}}

\title{Mechanism for nonlocal information flow from black holes}

\author[1]{Adithya Kandhadai} 
\author[1, 2]{Antony Valentini}
\affil[1]{\footnotesize Department of Physics and Astronomy, \protect\\Clemson University, Kinard Laboratory \protect\\Clemson, SC 29634-0978, USA.}
\affil[2]{\footnotesize Augustus College, \protect\\14 Augustus Road, \protect\\London SW19 6LN UK.}
\date{}

\begin{document}
\maketitle

\begin{abstract}
We show that quantum nonequilibrium (or deviations from the Born rule) can propagate nonlocally across space. Such phenomena are allowed in the de Broglie-Bohm pilot-wave formulation of quantum mechanics. We show that an entangled state can act as a channel whereby quantum nonequilibrium can be transferred nonlocally from one region to another without any classical interaction. This suggests a novel mechanism whereby information can escape from behind the classical event horizon of an evaporating black hole.
\end{abstract}

\section{Introduction}

In a hidden-variables theory that determines the outcomes of individual quantum events, Bell's theorem tells us that the underlying dynamics must be nonlocal.\footnote{With three notable caveats: it is assumed that there is no backwards causation, no `superdeterminism' for the apparatus settings, and that there is only one universe.}
This is the case, for example, in the pilot-wave theory of de Broglie and Bohm \cite{deB28, BV09, B52a, B52b, Holl93}, which associates definite trajectories with the wave function of an individual system. And yet, in such theories as they are usually presented, the statistics are local: the marginal distribution of measurement outcomes at one wing of an entangled state is unaffected by local operations performed at
the other distant wing. It is then not possible to utilise entangled states for superluminal signalling and a `peaceful co-existence' with relativity is assured. However, there are various reasons for considering the possibility of nonlocality at the level of statistics as well.

First of all, in a deterministic theory there is a clear conceptual
distinction between dynamical laws on the one hand and initial conditions on the other. In pilot-wave theory, for example, the initial conditions are given by the initial configuration $q(0)$ and the initial wave function $\psi(q,0)$. These evolve in time according to the laws of motion%
\begin{equation}
\frac{dq}{dt}=\frac{j}{\left\vert \psi\right\vert ^{2}}\label{deB1}%
\end{equation}
and%
\begin{equation}
i\frac{\partial\psi}{\partial t}=\hat{H}\psi\label{Sch1}%
\end{equation}
(the de Broglie guidance equation and the Schr\"{o}dinger equation
respectively), where $j=j[\psi]$ is the usual `quantum current' and $\hat{H}$ is an appropriate Hamiltonian operator. In applications to quantum mechanics, an assumption about initial conditions is added to these equations. The assumption is that, over an ensemble of systems with the same initial wave function $\psi(q,0)$, the initial configurations $q(0)$ have a distribution given by the Born rule:%
\begin{equation}
\rho(q,0)=\left\vert \psi(q,0)\right\vert ^{2}\ .\label{Born}%
\end{equation}
As is now well known, this assumption guarantees agreement with the empirical predictions of quantum mechanics \cite{B52a, B52b}. The outcomes of individual quantum measurements are in principle determined by the equations of motion (\ref{deB1}) and (\ref{Sch1}) alone. But in practice we are unable to control the individual initial conditions $q(0)$. By adding the statistical assumption (\ref{Born}), it may be shown that the ensemble distribution of outcomes of quantum measurements always agrees with the usual quantum predictions.
Similarly, for a general deterministic hidden-variables theory, where the initial conditions (associated with some initial state preparation at $t=0$) are usually denoted symbolically by $\lambda$, the outcomes $\omega$ of quantum measurements are presumed to be determined by some mapping \cite{AV07}%
\begin{equation}
\omega=\omega(M,\lambda)\ ,\label{omega}%
\end{equation}
where $M$ denotes the settings of the measurement apparatus. The form of the mapping (\ref{omega}) is usually not specified (unless one is discussing a particular model). But a specific theory will provide a specific such mapping, which in effect plays the role of an `equation of motion' -- mapping from
initial conditions to determined outcomes. Again, in principle, the outcomes of individual quantum measurements are determined by (\ref{omega}) alone. But in practice we are unable to control the individual initial conditions $\lambda$. We then introduce a statistical assumption, that over an ensemble of similar experiments the $\lambda$'s have a specific distribution%
\begin{equation}
\rho(\lambda)=\rho_{\mathrm{QT}}(\lambda)\label{Born2}%
\end{equation}
(for a given state preparation), where $\rho_{\mathrm{QT}}$ is chosen so that ensemble averages%
\begin{equation}
\left\langle \omega\right\rangle _{\mathrm{QT}}=\int d\lambda\ \rho
_{\mathrm{QT}}(\lambda)\omega(M,\lambda)
\end{equation}
agree with the averages predicted by quantum theory. Now the key
point is that equations such as (\ref{deB1}), (\ref{Sch1}) and (\ref{omega}) have the immutable character of dynamical laws: they are supposed to apply always and everywhere to any given individual system. In contrast, equations such as (\ref{Born}) and (\ref{Born2}) amount to statistical assumptions about
the initial conditions. It is usual to regard initial conditions as
contingencies, in the sense that there is no lawlike reason why they could not be different from what they happen to be. On this view, then, the equations (\ref{Born}) and (\ref{Born2}) do not have a lawlike status: they play an important role in explaining what we see, but if the underlying theory is
taken seriously we can and should entertain the possibility of more general distributions of initial conditions \cite{AV91a, AV91b}%
\begin{equation}
\rho(q,0)\neq\left\vert \psi(q,0)\right\vert ^{2}\label{non-Born}%
\end{equation}
and \cite{AV07}%
\begin{equation}
\rho(\lambda)\neq\rho_{\mathrm{QT}}(\lambda)\ .\label{non-Born2}%
\end{equation}

General distributions (\ref{non-Born}) and (\ref{non-Born2}) correspond to `quantum nonequilibrium', in contrast with the `quantum equilibrium' distributions (\ref{Born}) and (\ref{Born2}). This terminology is chosen by analogy with thermal physics, where both equilibrium and nonequilibrium distributions are possible (where classically the distributions are usually defined on phase space). In pilot-wave theory, the initial Born-rule
distribution (\ref{Born}) enjoys the property of `equivariance' under the equations of motion (\ref{deB1}) and (\ref{Sch1}): such a distribution evolves into a Born-rule distribution $\rho(q,t)=\left\vert \psi(q,t)\right\vert ^{2}$
at later times (see Section 2). Furthermore, as one might expect from the analogy with thermal physics, in appropriate circumstances initial nonequilibrium distributions relax towards equilibrium, as has been shown through extensive numerical simulations \cite{AV92, AV01, VW05, TRV12, SC12, ACV14}. Such relaxation may be quantified in terms of a subquantum $H$-function \cite{AV91a}%
\begin{equation}
H=\int dq\ \rho\ln(\rho/\left\vert \psi\right\vert ^{2})\ ,
\end{equation}
whose coarse-grained value is found to decay approximately exponentially with
time \cite{VW05, TRV12, ACV14}.

While there is as yet no experimental evidence for violations of the Born rule, the theoretical study of quantum nonequilibrium opens up a large domain of potentially new physics which might one day be observable (or at least testable). The new physics includes superluminal signalling (thereby requiring an underlying preferred foliation of spacetime) and violations of the uncertainty principle, as well as the breaking of other conventional quantum constraints (such as the indistinguishability of non-orthogonal states) \cite{AV91a, AV91b, AV92, AV02, AV04, AV09, PV06}. It is arguably natural to assume that the universe began in a state of quantum nonequilibrium, in which case it is possible that the Born rule was violated in the very early universe \cite{AV91a, AV91b, AV92, AV01, AV07, AV96, AV10, UV15}.\footnote{Some authors assert that initial nonequilibrium is intrinsically unlikely because it is `untypical' with respect to the equilibrium (Born-rule) measure \cite{DGZ92, Tum18}. Such arguments are, however, circular because they assume that the measure of typicality is indeed given by the equilibrium Born rule. Scientifically speaking, initial conditions in pilot-wave theory are ultimately an empirical question \cite{Allori}.} In the context of inflationary cosmology such violations could produce an observable imprint on the cosmic microwave background \cite{AV07, AV10, AVarx1, CV13, CV15, CV16} -- such as a large-scale power deficit, for which
there is tentative evidence \cite{SPV19}.

Another, less well-developed, line of inquiry concerns possible implications for the physics of black holes \cite{AVarx2, AV07}. This brings us to a second motivation for considering the possibility of nonlocality at the level of statistics. Ever since Hawking's seminal paper of 1976 \cite{Hawk76}, the paradox of information loss in black holes has remained controversial. The essential difficulty
identified by Hawking may be summarised as follows. Consider an isolated system described by an initial pure quantum state $\hat{\rho}_{i}=\left\vert
\Psi_{i}\right\rangle \left\langle \Psi_{i}\right\vert $ (perhaps defined in the remote past). Let the system undergo gravitational collapse to a black hole. The hole will emit thermal Hawking radiation and hence steadily lose mass (or `evaporate'). If we assume that the hole evaporates completely, we
are eventually left with just thermal radiation in some final \textit{mixed}
quantum state represented by a mixed density operator $\hat{\rho}_{f}$ (perhaps defined in the remote future). We then seem to have an isolated system that evolves from an initial pure state to a final mixed state -- contrary to the basic equations of quantum mechanics. The `information loss' arises from our inability to retrodict the initial pure state $\left\vert
\Psi_{i}\right\rangle $ from the final mixed state $\hat{\rho}_{f}$ (since the thermal radiation described by $\hat{\rho}_{f}$ appears to be independent of the details of $\left\vert \Psi_{i}\right\rangle $). There are numerous
caveats in the argument -- such as the assumption of complete evaporation -- which continue to cause controversy. Still, attempts to avoid what appears to be a breakdown of unitary evolution have been hugely influential, motivating for example the AdS/CFT correspondence \cite{Polch17}.

To understand the connection with nonlocality, we need to bear in mind a key aspect of the physics of Hawking radiation. The process is usually described in terms of a quantised field $\hat{\phi}$ propagating on a background classical spacetime representing a black hole. The field operator $\hat{\phi}$ is expanded in terms of `ingoing' and `outgoing' field modes, whose wave
vectors correspond to wave propagation into and out of the hole respectively. It is found that, in the natural vacuum state, the ingoing and outgoing field modes are entangled \cite{BD82}. Tracing over the ingoing modes then yields a mixed state $\hat{\rho}_{f}$ for the outgoing modes -- which is found to be
thermal. Heuristically, this essential physical point is often visualised in terms of pair creation near the event horizon, with one particle falling into the hole and its entangled partner propagating outwards. In any case, for our purposes the key point is that the outgoing degrees of freedom (making up the radiation in the exterior region) are entangled with ingoing degrees of
freedom hidden behind the horizon. It may then seem innocuous that the external state is mixed, since that state arises simply by tracing over inaccessible internal degrees of freedom. However, as the black hole evaporates away, there is eventually no interior region left to trace over: the final state $\hat{\rho}_{f}$ is then not merely an effective mixed state obtained by ignoring other degrees of freedom, instead $\hat{\rho}_{f}$ is the complete state and appears to be \textit{fundamentally} mixed.

More formally, the initial pure state $\left\vert\Psi_{i}\right\rangle $ will be defined on an initial spacelike hypersurface $\Sigma_{i}$. Once the black-hole horizon has formed and evaporation begins, the complete Hilbert space $\mathcal{H}$ of quantum states may be written as a product%
\begin{equation}
\mathcal{H}=\mathcal{H}_{\mathrm{int}}\otimes\mathcal{H}_{\mathrm{ext}}%
\end{equation}
over interior and exterior degrees of freedom. On a hypersurface $\Sigma$ that crosses the horizon, the quantum state is still $\left\vert \Psi_{i}\right\rangle $ (in the Heisenberg picture). But in the exterior region we have a mixed quantum state represented by a reduced density operator%
\begin{equation}
\hat{\rho}_{\mathrm{ext}}=\mathrm{Tr}_{\mathrm{int}}(|\Psi_{i}\rangle
\langle\Psi_{i}|)\ ,
\end{equation}
obtained by tracing over the interior degrees of freedom. After the black hole has completely evaporated, the mixed state $\hat{\rho}_{\mathrm{ext}}$ \textit{is} the final state $\hat{\rho}_{f}$ of the whole system (defined on a
final hypersurface $\Sigma_{f}$). We then seem to have generated a time evolution from an initial pure state to a final mixed state \cite{Wald94}.

Such issues have led many workers to seek a mechanism whereby information from behind the horizon can somehow make its way to the exterior region. At first sight, such a mechanism might seem to require superluminal signalling or at least a strong form of nonlocality. Indeed some workers, in particular
Giddings, have advocated some form of nonlocality as a solution \cite{Gidd06}. In the context of pilot-wave theory, it has been proposed that the entanglement between ingoing and outgoing degrees of freedom can act as a channel whereby information from behind the horizon propagates nonlocally to the exterior region, thereby potentially providing a mechanism whereby information loss may
be avoided \cite{AVarx2, AV07}. For this mechanism to work, it must be assumed that there exist internal degrees of freedom (behind the horizon) that are already in a state of quantum nonequilibrium, perhaps owing to new physics operating close to the singularity at the Planck scale \cite{AV10, AVPIRSA, AVprep}. To see
this, consider the ingoing and outgoing modes of the field $\hat{\phi}$. We assume that this field begins in quantum equilibrium. If the ingoing modes interact with internal nonequilibrium degrees of freedom then, as we shall see
in detail in this paper, the dynamics of pilot-wave theory shows that the outgoing modes can be knocked out of equilibrium -- by virtue of the entanglement operating nonlocally across the horizon. This is illustrated schematically in Figure 1.%

\begin{figure}[H]
\begin{center}
\includegraphics[height=2.8237in, width=5.0819in]
{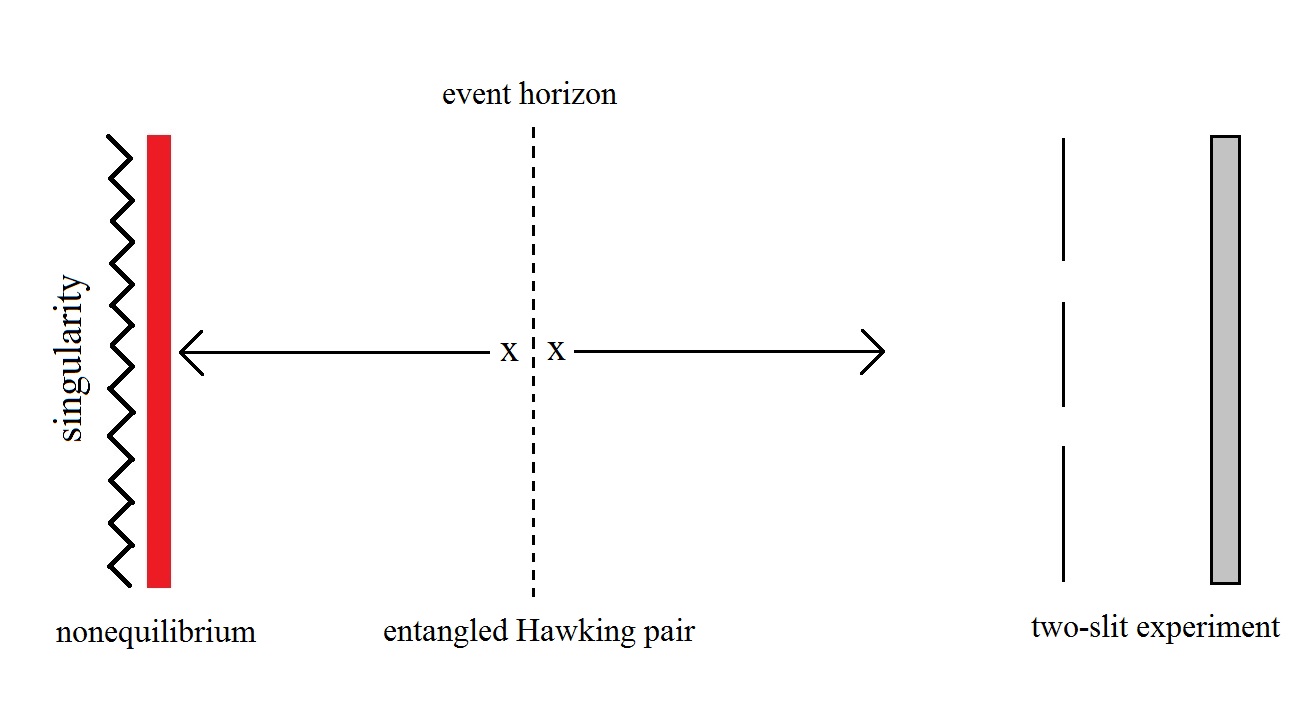}
\caption{Schematic mechanism for the nonlocal propagation of quantum
nonequilibrium from the interior of a black hole to the exterior region \cite{AV07, AVarx2}.}
\end{center}
\end{figure}

This is the proposed mechanism for the avoidance of black-hole information loss \cite{AVarx2, AV07}. Many details, however, remain to be studied. Presumably the form of the nonequilibrium distribution for the external particles will depend on the nonequilibrium distribution that was initially present in the interior, thereby providing a conduit for information flow from behind the horizon. While such details have not yet been studied, the proposal does have a clear qualitative implication: Hawking radiation is predicted to contain particles in a state of quantum nonequilibrium. This means that (some of) the radiated particles will violate the usual Born rule. An attempt was made to quantify this by the simple proposal or ansatz that the increase of von Neumann entropy
$S_{\mathrm{vonN}}=-\mathrm{Tr}(\hat{\rho}\ln\hat{\rho})$ associated with the pure-to-mixed transition ($\hat{\rho}_{i}\rightarrow\hat{\rho}_{f}$) be balanced by a decrease of the `subquantum' or hidden-variable entropy
$S_{\mathrm{hv}}=-H$ associated with nonequilibrium in the exterior region \cite{AVarx2, AV07}:%
\begin{equation}
\Delta S_{\mathrm{vonN}}+\Delta S_{\mathrm{hv}}=0\ .\label{Sconsn}%
\end{equation}
But the relation between the two kinds of entropy is poorly understood, so this remains only an \textit{ad hoc} hypothesis. In principle, however, these ideas could be tested. For example, should Hawking radiation be observed from evaporating primordial black holes (which in some scenarios are assumed to be
a significant component of dark matter \cite{CKS16}), the detected radiation could be tested for violations of the Born rule. In particular, we could test the radiation for violations of Malus' law for single-photon polarisation probabilities \cite{AV04} or simply test it for anomalies in two-slit interference (as indicated in Figure 1). Alternatively, since the proposal suggests that we may
expect to find quantum nonequilibrium at the Planck scale, it can also be tested with inflationary cosmology \cite{AV10}.

The purpose of this paper is to study the basic physics behind the proposed mechanism and to show, through examples, exactly how an entangled state can act as a conduit for the nonlocal propagation of quantum nonequilibrium. To illustrate the mechanism it will suffice to consider a simple and calculable model of low-energy entangled degrees of freedom. We show, both analytically and numerically, that indeed an equilibrium system can be thrown out of equilibrium if it is entangled with another equilibrium system that in turn interacts locally with a nonequilibrium system -- that, in other words,
nonequilibrium can propagate nonlocally across the light cone, in just the fashion required for the proposed solution to the black-hole information loss paradox.

In Section 2 we review some essential background in pilot-wave theory, in particular as applied to a scalar field in curved spacetime. In Section 3 we present our model and show analytically that, for the case of an impulsive interaction, quantum nonequilibrium does indeed propagate nonlocally. In
Section 4 we study a similar system numerically and show how it applies to field modes. In Section 5 we present our conclusions. 

\section{Pilot-wave theory and quantum equilibrium}
In pilot-wave theory, we can consider an ensemble of systems with an arbitrary initial distribution of configurations $q$ (in general unequal to the initial Born distribution $|\psi(q, t_i)|^2$). By construction, because each system in the distribution follows the velocity field $\dot{q}$ given by de Broglie's guidance equation (\ref{deB1}), it necessarily follows that the distribution $\rho$ evolves according to the continuity equation
\begin{equation}
\label{cont-eq}
\frac{\partial \rho}{\partial t} + \partial_q (\rho \dot{q}) = 0.
\end{equation}
Now the Schrödinger equation implies that the Born distribution $|\psi|^2$ obeys the same continuity equation with the same velocity field. It follows that if a system is initially in quantum equilibrium, that is if $\rho(q, t_i)=|\psi(q, t_i)|^2$, then $\rho$ and $|\psi|^2$ will evolve identically and hence the Born rule will be satisfied at all future times.

In the following subsections we summarize pilot-wave theory for (spinless) particles and fields. We present pilot-wave theory on a curved spacetime background, and we show that field theory may be conveniently recast in terms of degrees of freedom in Fourier space. This sets the scene for the calculations that follow.

\subsection{Pilot-wave theory of low-energy particles}

For a single particle, the configuration $q$ is simply the particle position $\textbf{x}(t)$ and the wave function is in effect a complex-valued field in 3-space. If the Hamiltonian is of the standard form with kinetic and potential terms,
\begin{equation}
\label{1-particle Hamiltonian}
\hat{H}=-\frac{1}{2m} \nabla^2 + V(\textbf{x}),
\end{equation}
the de Broglie velocity field is proportional to the gradient of the phase $S$ of the wave function $\psi = |\psi|e^{iS}$. The guidance equation (\ref{deB1}) then takes the form
\begin{equation}
\label{1-particle velocity field}
\frac{d \textbf{x}}{d t} = \frac{1}{m}\nabla S.
\end{equation}
More generally, for a system of $n$ particles with a Hamiltonian of the form 
\begin{equation}
\label{n-particle Hamiltonian}
\hat{H} = \sum_{i=1}^{n} -\frac{1}{2m_i}\nabla_i^2 + V(\textbf{x}_i),
\end{equation}
the velocity field for the $i^{th}$ particle is given by 
\begin{equation}
\label{n-particle velocity field}
\frac{d \textbf{x}_i}{d t} = \frac{1}{m_i} \nabla_i S,
\end{equation}
where $S$ is the phase of the total wave function.

It is sometimes convenient to decompose the Schrödinger equation into its real and imaginary parts, working with the wave function amplitude $|\psi|$ and the phase $S$. For the system of $n$ particles this yields the continuity equation
\begin{equation}
\label{n-particle continuity equation}
\frac{\partial |\psi|^2}{\partial t} + \sum_{i=1}^{n}\nabla_i \cdot \left(|\psi|^2\frac{1}{m_i} \nabla_i S\right) = 0    
\end{equation}
and the modified Hamilton-Jacobi equation
\begin{equation}
\label{Hamilton-Jacobi}
\frac{\partial S}{\partial t} + \sum_{i=1}^{n} \frac{1}{2m_i}(\nabla_i S)^2 + Q + V = 0    
\end{equation}
where 
\begin{equation}
\label{quantum potential}
Q = -\sum_{i=1}^{n}\frac{1}{2m_i}\frac{1}{|\psi|}\nabla_i^2 |\psi|
\end{equation}
is the `quantum potential' \cite{B52a, B52b}.

\subsection{Pilot-wave field theory on curved spacetime}

The proper setting for a discussion of black-hole information loss is of course quantum field theory on a curved spacetime background. We will not be using this formalism in the calculations reported here, which are carried out for a simplified model in Minkowski spacetime. But even so it is important to show the connection with a more complete model on curved spacetime. We may restrict ourselves to a scalar field.

It is in fact straightforward to write down pilot-wave theory for a scalar field propagating on a classical curved spacetime background \cite{AVarx2}. Assuming as usual that the spacetime is globally hyperbolic, it can be foliated (generally nonuniquely) by spacelike hypersurfaces $\Sigma(t)$ labelled by a global time parameter $t$. The spacetime line element $d\tau^{2}=\,^{(4)}g_{\mu\nu}dx^{\mu}dx^{\nu}$ with 4-metric $^{(4)}g_{\mu\nu}$ can
then be decomposed as \cite{ADM, HE73}%
\begin{equation}
d\tau^{2}=(N^{2}-N_{i}N^{i})dt^{2}-2N_{i}dx^{i}dt-g_{ij}dx^{i}dx^{j}%
\ ,\label{ADM}%
\end{equation}
where $N$ is the lapse function, $N^{i}$ is the shift vector, and $g_{ij}$ is the 3-metric on $\Sigma(t)$. We may set $N^{i}=0$ (provided the lines $x^{i}=\mathrm{const}.$ do not meet singularities).

For a massless real scalar field $\phi$ with Lagrangian density%
\begin{equation}
\mathcal{L}=\frac{1}{2}\sqrt{-\,^{(4)}g}\,^{(4)}g^{\mu\nu}\partial_{\mu}%
\phi\partial_{\nu}\phi
\end{equation}
the canonical momentum density reads $\pi=\partial\mathcal{L}/\partial\dot{\phi}=(\sqrt{g}/N)\dot{\phi}$ (where $^{(4)}g=\det g_{\mu\nu}$ and $g=\det g_{ij}$) and we have a classical Hamiltonian%
\begin{equation}
H=\int d^{3}x\;\frac{1}{2}N\sqrt{g}\left(  \frac{1}{g}\pi^{2}+g^{ij}%
\partial_{i}\phi\partial_{j}\phi\right)  \ .
\end{equation}
The wave functional $\Psi\lbrack\phi,t]$ then obeys the Schr\"{o}dinger equation%
\begin{equation}
i\frac{\partial\Psi}{\partial t}=\int d^{3}x\;\frac{1}{2}N\sqrt{g}\left(
-\frac{1}{g}\frac{\delta^{2}}{\delta\phi^{2}}+g^{ij}\partial_{i}\phi
\partial_{j}\phi\right)  \Psi\ .\label{Sch2}%
\end{equation}
From this we obtain a continuity equation%
\begin{equation}
\frac{\partial\left\vert \Psi\right\vert ^{2}}{\partial t}+\int d^{3}%
x\;\frac{\delta}{\delta\phi}\left(  \left\vert \Psi\right\vert ^{2}\frac
{N}{\sqrt{g}}\frac{\delta S}{\delta\phi}\right)  =0\label{cont2}%
\end{equation}
with a current $j=\left\vert \Psi\right\vert ^{2}(N/\sqrt{g})\delta S/\delta\phi$ and a de Broglie velocity field%
\begin{equation}
\frac{\partial\phi}{\partial t}=\frac{N}{\sqrt{g}}\frac{\delta S}{\delta\phi
}\ ,\label{deB2}%
\end{equation}
where as usual $\Psi=\left\vert \Psi\right\vert e^{iS}$ \cite{AVarx2}.

The time derivative (\ref{deB2}) of the field at a point $x^{i}$ on $\Sigma(t)$ depends instantaneously (with respect to $t$) on the field at distant points $(x^{\prime})^{i}\neq x^{i}$ -- provided $\Psi$ is entangled with respect to the field values at those points. Physical consistency is ensured if we assume that the theory is constructed with a chosen preferred foliation (associated with some lapse function $N(x^{i},t)$) \cite{AV08a}.

An arbitrary distribution $P[\phi,t]$ will necessarily satisfy the same continuity equation:%
\begin{equation}
\frac{\partial P}{\partial t}+\int d^{3}x\;\frac{\delta}{\delta\phi}\left(
P\frac{N}{\sqrt{g}}\frac{\delta S}{\delta\phi}\right)  =0\ .\label{cont2'}%
\end{equation}
It follows that if $P=\left\vert \Psi\right\vert ^{2}$ holds at some initial time it will hold at all times.

For the purposes of the rest of this paper, it will suffice to consider (for simplicity) a scalar field in Minkowski spacetime and to show how quantum nonequilibrium can propagate nonlocally across the lightcone. This suffices to illustrate the physical mechanism, and moreover provides us with a more tractable model.

\subsection{Pilot-wave field theory and harmonic oscillators}

Consider, then, a real massless scalar field $\phi$ on Minkowski spacetime. The Lagrangian density is now given by $\mathscr{L} = \frac{1}{2} \sqrt{-g} g^{\mu \nu} \partial_{\mu} \phi \partial_{\nu} \phi$. In a given frame with time parameter $t$, the Lagrangian can be written as 
\begin{equation}
\label{scalar field Lagrangian}
L = \frac{1}{2} \int d^3 \textbf{x} \left( \dot{\phi}^2 - (\nabla \phi)^2 \right).
\end{equation}
It will prove convenient to rewrite the theory in terms of the Fourier components 

$$ \phi_{\textbf{k}}(t) = \frac{1}{(2 \pi)^{\frac{3}{2}}} \int d^3 \textbf{x} \: \phi(\textbf{x}, t) e^{-i \textbf{k}.\textbf{x}}.  $$
These can be decomposed into real and imaginary parts,  
\begin{equation}
\label{Fourier components decomposition}
\phi_{\textbf{k}} = \frac{\sqrt{V}}{(2 \pi)^{\frac{3}{2}}}(q_{\textbf{k}1} + iq_{\textbf{k}2}),
\end{equation}
giving two real degrees of freedom, $q_{\textbf{k}1}$ and $q_{\textbf{k}2}$ for each mode (here $V$ is a normalization volume).
The Lagrangian becomes
\begin{equation}
\label{Lagrangian rewritten}
L = \frac{1}{2} \sum_{\textbf{k} r} \left(\dot{q}_{\textbf{k}r}^2 - k^2 q_{\textbf{k}r}^2 \right).
\end{equation}
We then have canonical momenta $\pi_{\textbf{k}r} = \frac{\partial L}{\partial \dot{q}_{\textbf{k}r}} = \dot{q}_{\textbf{k}r}$ and a classical Hamiltonian
\begin{equation}
\label{scalar field Hamiltonian}
H = \frac{1}{2} \sum_{\textbf{k} r} (\pi_{\textbf{k} r}^2 + k^2 q_{\textbf{k} r}^2).
\end{equation}
Hence $H = \sum_{\textbf{k} r} H_{\textbf{k} r}$, where $H_{\textbf{k} r}$ is equivalent to the Hamiltonian of a one-dimensional harmonic oscillator with mass $m = 1$ and angular frequency $\omega = k$.

We now focus on a decoupled {\textendash} that is, unentangled {\textendash} mode of wave number $\textbf{k}$. The total wave functional $\Psi$ takes the form $\Psi = \psi_{\textbf{k}}(q_{\textbf{k}1}, q_{\textbf{k}2}, t)\chi$ where $\chi$ has no dependence on degrees of freedom for the mode $\textbf{k}$. The mode $\textbf{k}$ then has its own independent dynamics. The decoupled wave function $\psi_{\textbf{k}}$ obeys the Schrödinger equation
\begin{equation}
\label{field mode Schrodinger equation}
i \frac{\partial \psi_{\textbf{k}}}{\partial t} = \sum_{r = 1}^{2} \left( -\frac{1}{2}\frac{\partial ^2}{\partial q_{\textbf{k} r}^2} + \frac{1}{2}k^2 q_{\textbf{k} r}^2 \right) \psi_{\textbf{k}}.
\end{equation}
This has the same form as the Schrödinger equation for a two-dimensional harmonic oscillator with degrees of freedom $q_{\textbf{k}1}$ and $q_{\textbf{k}2}$ and with unit mass and angular frequency $k$.

The de Broglie guidance equations for the degrees of freedom $q_{\textbf{k}1}$ and $q_{\textbf{k}2}$ are simply
\begin{equation}
\label{field mode guidance equation}
\frac{dq_{\textbf{k}r}}{dt} = \frac{\partial s_\textbf{k}}{\partial q_{\textbf{k}r}} \quad (r=1, 2),
\end{equation}
where $s_{\textbf{k}}$ is the phase of $\psi_{\textbf{k}}$. These are again of the same form as the guidance equations for a two-dimensional harmonic oscillator of unit mass. Thus a decoupled field mode of wave number $\textbf{k}$ can be studied in terms of the analogous two-dimensional harmonic oscillator. 

If we now assume that the decoupled wave function $\psi_{\textbf{k}} (q_{\textbf{k}1}, q_{\textbf{k}2}, t)$ is further separable as $\psi_{\textbf{k}} = \psi_{\textbf{k}1} (q_{\textbf{k}1}, t) \psi_{\textbf{k}2} (q_{\textbf{k}2}, t) $, each of the degrees of freedom $q_{\textbf{k}r}$ behaves like the position of a one-dimensional harmonic oscillator of unit mass and angular frequency $k$. In this way, one-dimensional harmonic oscillators can model the degrees of freedom of a quantum field under particular conditions. This correspondence is used extensively in our calculations below, which feature particles moving in one dimension. The results are equally applicable to quantum field theory for decoupled field modes.

\section{Nonlocal propagation of quantum nonequilibrium}

We now study a simple model that illustrates the nonlocal propagation of quantum nonequilibrium.

\subsection{Model summary}

The model involves a pair of entangled particles A and B, which are initially in quantum equilibrium, and a third particle C, which is initially in nonequilibrium. We investigate whether an interaction of B with C can pull A out of equilibrium, even when A does not interact directly with B or C. The basic setup is illustrated in Figure 2 below.

\begin{figure}[H]
\centering
\includegraphics[width=10cm, height=3cm]{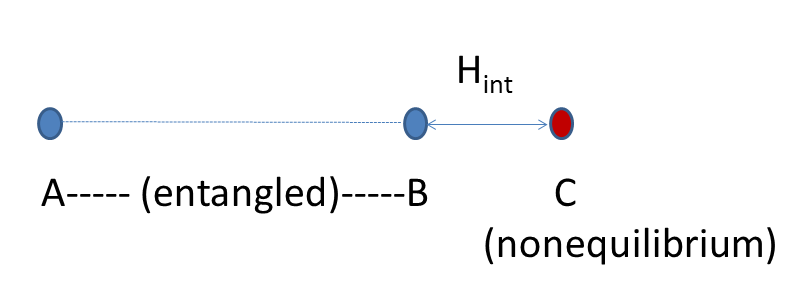}
\caption{An illustration of the three-particle problem.}
\end{figure}

We first need to define the concept of nonequilibrium for a particle which is entangled with other particles. In such a case the particle is said to be in non-equilibrium if its marginal distribution differs from the equilibrium marginal distribution (where the latter is found by taking the squared-amplitude of the total wavefunction and integrating over the other degrees of freedom). 

In our model, each particle has just one degree of freedom {\textendash} the respective position on the $x$-axis $x_A$, $x_B$ and $x_C$. This suffices to capture the essential features of the problem. The joint wavefunction is denoted by $\Psi(x_A, x_B, x_C, t) = R e^{iS}$ (where it is useful to introduce the amplitude and phase $R$ and $S$, respectively). The initial wavefunction may be written as 
\begin{equation}
\label{initial 3-particle wave function}
\Psi_0(x_A, x_B, x_C) = \psi_0(x_A, x_B)\chi_0(x_C) = R_0e^{iS_0}. 
\end{equation} 
The equilibrium marginal distribution of particle A is given by
\begin{equation}
\label{marg_eq_A}
\rho_A^{\textrm{eq}}(x_A, t) = \int_{-\infty}^{\infty}dx_C \int_{-\infty}^{\infty}dx_B |\Psi(x_A, x_B, x_C, t)|^2.
\end{equation}
The joint probability distribution of all three particles is denoted by $\rho(x_A, x_B, x_C, t)$. The initial joint distribution may be written as
\begin{equation}
\label{initial 3-particle distribution}
\rho_0(x_A, x_B, x_C) = |\psi_0(x_A, x_B)|^2q_0(x_C).
\end{equation}
The marginal distribution of A is then given by
\begin{equation}
\label{marg_A}
\rho_A(x_A, t) = \int_{-\infty}^{\infty}dx_C \int_{-\infty}^{\infty}dx_B \rho(x_A, x_B, x_C, t).
\end{equation}
Since particle C is initially in nonequilibrium, $q_0(x_C)$ differs from $|\chi_0(x_C)|^2$. 

We now need to pick a suitable interaction Hamiltonian. It is clear that if the Hamiltonian generates a velocity field for A that is identically zero, the marginal distribution for A will not change from its initial value, and A will hence stay in equilibrium. This can be avoided by including the kinetic energy of particle A in the Hamiltonian.

We construct an example where the only interaction is between particles B and C, with A evolving according to its free particle Hamiltonian $H_A = p_A^2/2m$. In addition to their own free particle Hamiltonians $H_B = p_B^2/2m$ and $H_C = p_C^2/2m$, B and C have an interaction term $H_{intBC}$.  Thus the total Hamiltonian $H_{ABC}$ is given by
\begin{equation}
\label{3-particle Hamiltonian}
H_{ABC} = \frac{p_A^2}{2m} + \frac{p_B^2}{2m} + \frac{p_C^2}{2m} + H_{intBC}.
\end{equation}

In order to proceed we first derive some useful general results. 

The Born distribution $|\Psi|^2$ obeys a continuity equation with velocity fields which we denote by $v_A$, $v_B$ and $v_C$:
\begin{equation}
\label{3-particle general continuity equation for Born distribution}
\frac{\partial}{\partial t}(|\Psi|^2) + \frac{\partial}{\partial x_A} (|\Psi|^2 v_A) + \frac{\partial}{\partial x_B} (|\Psi|^2 v_B) + \frac{\partial}{\partial x_C} (|\Psi|^2 v_C) = 0.
\end{equation}
Integrating over $x_B$ and $x_C$ implies that the marginal for A satisfies
$$ \frac{\partial}{\partial t}(\rho_A^{\textrm{eq}}) = -\frac{\partial}{\partial x_A} \left( \int_{-\infty}^{\infty}dx_C \int_{-\infty}^{\infty}dx_B |\Psi|^2.v_A \right). $$
The other terms vanish, because they become boundary terms with a factor of $|\Psi|^2$, which vanishes at both infinities.

The joint distribution $\rho$ obeys a continuity equation with the same velocity fields as in equation (\ref{3-particle general continuity equation for Born distribution}),
\begin{equation}
\label{3-particle general continuity equation for actual distribution}
\frac{\partial}{\partial t}(\rho) + \frac{\partial}{\partial x_A} (\rho v_A) + \frac{\partial}{\partial x_B} (\rho v_B) + \frac{\partial}{\partial x_C} (\rho v_C) = 0.   
\end{equation}

One can hence write an equation for the marginal $\rho_A$:

$$ \frac{\partial}{\partial t}(\rho_A) = -\frac{\partial}{\partial x_A} \left( \int_{-\infty}^{\infty}dx_C \int_{-\infty}^{\infty}dx_B \rho v_A \right). $$
We may then write
\begin{equation}
\label{first time derivative of difference between marginals}
\frac{\partial}{\partial t}(\rho_A - \rho_A^{\textrm{eq}}) = -\frac{\partial}{\partial x_A} \left( \int_{-\infty}^{\infty}dx_C \int_{-\infty}^{\infty}dx_B (\rho - |\Psi|^2)v_A \right).
\end{equation}

Now, it would suffice to construct an example where the expression (\ref{first time derivative of difference between marginals}) has a non-zero value after particles B and C interact, showing that the marginal distribution for particle A and its equilibrium counterpart are in the process of diverging. However, for the simplest analytically solvable cases, this quantity turns out to be zero immediately after the interaction. We therefore go one step further, finding a general expression for the second time derivative
$$ \frac{\partial^2}{\partial t^2}(\rho_A - \rho_A^{\textrm{eq}}) = -\frac{\partial}{\partial x_A} \left( \int_{-\infty}^{\infty} dx_C \int_{-\infty}^{\infty} dx_B \frac{\partial}{\partial t} \left ((\rho - |\Psi|^2)v_A \right) \right), $$
and seeking an example where this quantity does not vanish after the interaction between B and C.

Since $ v_A = \frac{1}{m_A} \frac{\partial S}{\partial x_A} $, and taking the three particles to be of equal mass $m$, we have

\begin{equation}
\label{second time derivative of difference between marginals}
\frac{\partial^2}{\partial t^2}(\rho_A - \rho_A^{\textrm{eq}}) = - \frac{1}{m}\frac{\partial}{\partial x_A} \left( \int_{-\infty}^{\infty} dx_C \int_{-\infty}^{\infty} dx_B \left( \frac{\partial}{\partial t} (\rho - |\Psi|^2).\frac{\partial S}{\partial x_A} + (\rho - |\Psi|^2).\frac{\partial}{\partial x_A}\left( \frac{\partial S}{\partial t} \right) \right) \right).
\end{equation}
We now need expressions for $\partial(\rho - |\Psi|^2)/\partial t$ and $\partial S/\partial t$. The former can be calculated from the continuity equation and the latter from the Schrödinger equation. However rather than calculating these quantities for a general Hamiltonian it is easier to take a specific example. 

\subsection{Case of an impulsive interaction Hamiltonian}

To show particle A's departure from quantum equilibrium, we proceed as follows. Consider an impulsive interaction between B and C that acts at $t = 0$, causing an instantaneous change to the joint state of the three particles. The dynamics due to the free Hamiltonian can be neglected during this period. If the expression in equation (\ref{second time derivative of difference between marginals}) becomes non-zero after the impulsive interaction, then we know that the marginal distribution $\rho_A$ and its equilibrium counterpart $\rho_A^{\textrm{eq}}$ cannot remain identical. In other words, the dynamics of the marginal distribution for A and its equilibrium counterpart are not the same, resulting in their divergence. 

We now choose the following simple interaction between B and C:
\begin{equation}
\label{choice of interaction Hamiltonian}
H_{intBC} = \alpha x_C p_B \delta (t).
\end{equation}
It is difficult to calculate (\ref{second time derivative of difference between marginals}) analytically, but if we treat $\alpha$ as a perturbative expansion parameter then (\ref{second time derivative of difference between marginals}) can be shown to be non-zero to first order in $\alpha$.

Let $\Psi^\prime (x_A, x_B, x_C) = R^\prime e^{iS^\prime}$ denote the joint wave function of the three particles immediately after the impulsive interaction. Integrating the Schrödinger equation
$$ i\frac{\partial \Psi}{\partial t} = -i\alpha \delta(t) x_C \frac{\partial \Psi}{\partial x_B}. $$
across $t = 0$ we readily find
\begin{equation}
\label{perturbative wave function immediately after interaction}
\Psi^\prime = \Psi_0 - \alpha x_C \frac{\partial \Psi_0}{\partial x_B}.
\end{equation}
Using this expession for $\Psi^\prime$ to calculate $R^\prime$ and $S^\prime$ for later use, we find up to lowest order in $\alpha$:
\begin{equation}
\label{amplitude immediately after interaction}
R^\prime = R_0 - \alpha x_C \frac{\partial R_0}{\partial x_B}.
\end{equation}
\begin{equation}
\label{phase immediately after interaction}
S^\prime = S_0 - \alpha x_C \frac{\partial S_0}{\partial x_B}.
\end{equation}
Similarly, integrating the continuity equations (\ref{3-particle general continuity equation for Born distribution}) and (\ref{3-particle general continuity equation for actual distribution}) across $t = 0$, we obtain the first-order corrections to $|\Psi|^2$ and $\rho$:
\begin{equation} 
\label{Born distribution immediately after interaction}
|\Psi^\prime|^2 = |\Psi_0|^2 - \alpha x_C \frac{\partial |\Psi_0|^2}{\partial x_B},
\end{equation}

\begin{equation}
\label{actual distribution immediately after interaction}
\rho^\prime = \rho_0 -  \alpha x_C \frac{\partial \rho_0}{\partial x_B}.
\end{equation}

After the impulsive interaction at $t = 0$, the Hamiltonian is taken to be simply equal to the free Hamiltonian for the three particles. We can now calculate the expressions $\frac{\partial}{\partial t}(\rho - |\Psi|^2)$ and $\frac{\partial S}{\partial t}$ for this Hamiltonian \footnote{The de Broglie-Bohm velocity fields which appear in the continuity equation are very much dependent on the Hamiltonian. A procedure for deriving the velocity fields for a general Hamiltonian was given by Struyve and Valentini \cite{SV09}.}, which follow from the continuity equations (\ref{3-particle general continuity equation for Born distribution}) and (\ref{3-particle general continuity equation for actual distribution}) and the modified Hamilton-Jacobi equation (see (\ref{Hamilton-Jacobi})) respectively:

\begin{equation}
\label{rate of change of actual minus Born}
\begin{aligned}
\frac{\partial}{\partial t}(\rho - |\Psi|^2) = -\frac{1}{m} \biggl( \frac{\partial}{\partial x_A} \left( (\rho - |\Psi|^2) \frac{\partial S}{\partial x_A} \right) + \frac{\partial}{\partial x_B} \left( (\rho - |\Psi|^2) \frac{\partial S}{\partial x_B} \right) \\ + \frac{\partial}{\partial x_C} \left( (\rho - |\Psi|^2) \frac{\partial S}{\partial x_C} \right) \biggr),
\end{aligned}
\end{equation}

\begin{equation}
\label{rate of change of phase}
\frac{\partial S}{\partial t} = \frac{1}{2} \biggl( \frac{1}{R}\frac{\partial^2 R}{\partial x_A^2} - \left(\frac{\partial S}{\partial x_A}\right)^2 + \frac{1}{R}\frac{\partial^2 R}{\partial x_B^2} - \left(\frac{\partial S}{\partial x_B}\right)^2 + \frac{1}{R}\frac{\partial^2 R}{\partial x_C^2} - \left(\frac{\partial S}{\partial x_C}\right)^2 \biggr).
\end{equation}
Combining equation (\ref{second time derivative of difference between marginals}) with equations (\ref{rate of change of actual minus Born}) and (\ref{rate of change of phase}), we have all we need to show that the marginal for particle A diverges from the equilibrium marginal after the impulsive interaction (for a specific example). All we need to do is calculate the quantity $\frac{\partial^2}{\partial t^2}(\rho_A - \rho_A^{\textrm{eq}})$ immediately after the impulsive interaction using expressions for $R^\prime$, $S^\prime$ and $\rho^\prime$.

\subsection{A numerical example}
After close inspection of equations (\ref{second time derivative of difference between marginals}), (\ref{rate of change of actual minus Born}) and (\ref{rate of change of phase}), we choose the following initial conditions immediately before the interaction in order to produce a non-null result with fairly simple expressions:

$$ R_0 = \frac{1}{\pi^\frac{3}{4}} e^{-\frac{(x_A^2 + x_B^2 + x_C^2)}{2}}, $$
$$ S_0 = x_A( x_B + x_B^2), $$ 
$$ \rho_0 = \frac{1}{\pi^\frac{3}{2}} e^{-(x_A^2 + x_B^2 + (x_C-d)^2)}. $$

Immediately after the impulsive interaction the corresponding quantities up to first order in $\alpha$ are

$$ R^\prime = \frac{1}{\pi^\frac{3}{4}}(1 + \alpha x_B x_C) e^{-\frac{(x_A^2 + x_B^2 + x_C^2)}{2}}, $$
$$ S^\prime = x_A(x_B - \alpha x_C + x_B^2 - 2\alpha x_B x_C), $$
$$ \rho^\prime = \frac{1}{\pi^\frac{3}{2}}(1 + 2\alpha x_B x_C) e^{-(x_A^2 + x_B^2 + (x_C - d)^2)}. $$
Using these quantities in equations (22), (29) and (30), we have the result (up to first order in $\alpha$)
\begin{equation}
\label{definition of f(x_A)}
\frac{\partial^2}{\partial t^2}(\rho_A - \rho_A^{\textrm{eq}}) = f(x_A) = \frac{6 \alpha d}{\sqrt{\pi} m}e^{-x_A^2}(1 - 2x_A^2).
\end{equation}
The first derivative $\frac{\partial}{\partial t}(\rho_A - \rho_A^{\textrm{eq}})$ is identically zero, and $f(x_A)$ also integrates to zero. The latter is as required for $\rho_A$ and $\rho_A^{eq}$ to remain normalized.

Taking values $\alpha = 0.025$, $d = 1$ and $m = 1$ (in appropriate units), for illustration we may plot the function $f(x_A)$ (see Figure 3).

\begin{figure}[H]
\centering
\includegraphics[width=7cm, height=5cm]{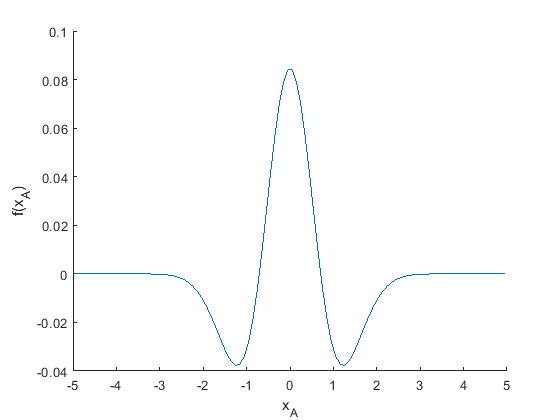}
\caption{A plot of the function $f(x_A)$ for $\alpha = 0.025$, $d = 1$ and $m = 1$.}
\end{figure}

Thus we see that, for this example, the marginal at A does indeed evolve away from equilibrium as claimed, even though there is no conventional interaction between A (which is initially in equilibrium) and C (which is initially in nonequilibrium).

\section{Simulations}

In the previous section, we gave an analytical example of our desired phenomenon. It is useful to consider a numerical simulation of the effect, with a view to illustrating the divergence between the marginal distribution for particle A and its equilibrium counterpart. 

\subsection{Setup}

We simulate an instance of the three-particle problem with the following initial conditions. The particles are harmonic oscillators with identical masses and frequencies (all set to unity). The initial state of A and B is an equal superposition of the $|01>$ and $|10>$ states of the equivalent two-dimensional harmonic oscillator, with a relative phase factor in order to avoid inconvenient nodal lines. Particles A and B start in equilibrium, so their joint initial distribution is the Born distribution corresponding to their initial wave function. The initial state of C is the ground state of the harmonic oscillator, and its initial distribution is the same Gaussian shifted by two units to the right. (In the equations below, $\phi_n(x)$ refers to energy eigenstates of the one-dimensional harmonic oscillator, with $\phi_0(x)$ being the ground state, $\phi_1(x)$ the first excited state, and so on, where these are all real functions.) Thus we have

$$ \psi_{AB}(x_A, x_B, 0) = \frac{1}{\sqrt{2}} \left[ \phi_0(x_A)\phi_1(x_B) + i\phi_1(x_A)\phi_0(x_B) \right], $$
$$ \rho_{AB}(x_A, x_B, 0) = \frac{1}{2} \left[ |\phi_0(x_A)|^2|\phi_1(x_B)|^2 + |\phi_1(x_A)|^2|\phi_0(x_B)|^2 \right], $$
$$ \psi_C(x_C, 0) = \phi_0(x_C), $$
$$ \rho_C(x_C, 0) = |\phi_0(x_C - 2)|^2. $$
The expression for $\rho_{AB}(0)$ does not feature cross-terms because, as mentioned, $\phi_0$ and $\phi_1$ are real functions.

The interaction Hamiltonian is the same as in equation (\ref{choice of interaction Hamiltonian}).

Figure 4 illustrates the initial conditions of the system just before the interaction takes place.

\begin{figure}[H]
\centering
\includegraphics[width=10cm, height=4.5cm]{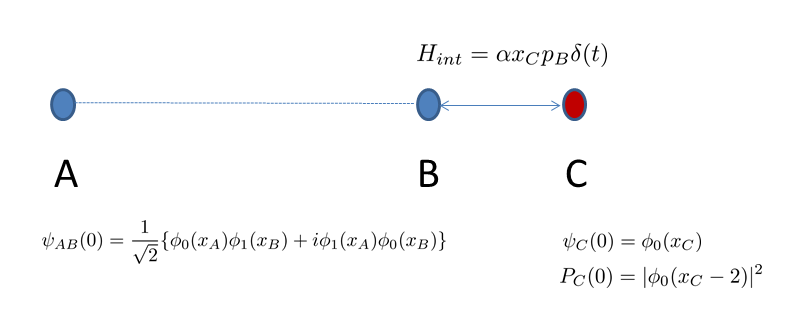}
\caption{The initial conditions for the three-particle problem as used in our numerical simulation. Particles A and B begin in equilibrium, so their initial distribution is simply $|\psi_{AB}(0)|^2$.}
\end{figure}

The wave function immediately after the interaction, which we will denote by $\Psi(0+)$, is obtained by integrating the Schrödinger equation as $\Psi(t) = e^{-i\hat{H}t} \Psi(0)$. Putting in the Hamiltonian from equation (\ref{choice of interaction Hamiltonian}) and expanding the exponential yields a Taylor series for $\Psi(0+)$ which may then be expressed in the closed form
\begin{equation}
\label{expanded wave function immediately after interaction}
\Psi(x_A, x_B, x_C, 0+) = \frac{1}{\sqrt{2}}(\phi_0(x_A)\phi_1(x_B - \alpha x_C) + i\phi_1(x_A)\phi_0(x_B - \alpha x_C))\phi_0(x_C).
\end{equation}
Using the continuity equation (\ref{3-particle general continuity equation for actual distribution}) for the actual distribution $\rho$ and integrating across $t = 0$ we obtain
\begin{equation}
\label{expanded distribution immediately after interaction}
\rho(x_A, x_B, x_C, 0+) = \frac{1}{2}\left(|\phi_0(x_A)\phi_1(x_B - \alpha x_C)|^2 + |\phi_1(x_A)\phi_0(x_B - \alpha x_C)|^2\right)|\phi_0(x_C-2)|^2.
\end{equation}

It can be seen from equations (\ref{expanded wave function immediately after interaction}) and (\ref{expanded distribution immediately after interaction}) that particles B and C are now out of equilibrium, while particle A is not (yet). Now it remains to evolve the particles under their Hamiltonian (a sum of three harmonic-oscillator Hamiltonians) and to see if the marginal distribution for A diverges from its equilibrium counterpart.

In order to calculate the trajectories of the particles, we need an expression for the wave function at any time $t$. It is difficult to obtain the exact wave function, which is given by the action of the time evolution operator $e^{-i\hat{H}t}$ on the wave function immediately after the interaction (equation (\ref{expanded wave function immediately after interaction})). However, since $\alpha$ is small, a good approximation for $\Psi(0+)$ may be obtained by expanding equation (\ref{expanded wave function immediately after interaction}) in powers of $\alpha$ and truncating the expansion at terms of order $\alpha^2$. In this way we obtain
\begin{equation}
\label{expanded perturbative wave function immediately after interaction}
\begin{aligned}
\Psi(0+) = \frac{1}{\sqrt{2}} & \biggl( \phi_0(x_A) \left( \phi_1(x_B) - \alpha x_C \phi_1^{\prime}(x_B) + \frac{\alpha^2 x_C^2}{2} \phi_1^{\prime \prime}(x_B) \right) \\ 
& + i\phi_1(x_A) \left( \phi_0(x_B) - \alpha x_C \phi_0^{\prime}(x_B) + \frac{\alpha^2 x_C^2}{2} \phi_0^{\prime \prime}(x_B) \right) \biggr) \phi_0(x_C).  \\
\end{aligned}
\end{equation}
We now rewrite this expansion as a superposition of eigenstates for the joint Hamiltonian of the three particles. Each eigenstate is represented by $\phi_j(x_A) \phi_k(x_B) \phi_l(x_C)$ and abbreviated as $\phi_{jkl}$, with $j$, $k$ and $l$ being non-negative integers. The energy eigenvalue corresponding to state $\phi_{jkl}$ is $E_{jkl} = j + k + l + \frac{3}{2}$. The following results for the one-particle eigenstates can be verified using the explicit functional forms of $\phi_n(x)$:

$$ \phi_0^{\prime}(x) = -\frac{1}{\sqrt{2}}\phi_1(x), \qquad \phi_0^{\prime \prime}(x) = \frac{1}{\sqrt{2}}\phi_2(x) - \frac{1}{2}\phi_0(x), $$

$$ \phi_1^{\prime}(x) = \frac{1}{\sqrt{2}}\phi_0(x) - \phi_2(x), \qquad \phi_1^{\prime \prime}(x) = \frac{\sqrt{3}}{2} \phi_3(x) - \frac{3}{2}\phi_1(x), $$

$$ x\phi_0(x) = \frac{1}{\sqrt{2}} \phi_1(x), \qquad x^2\phi_0(x) = \frac{1}{\sqrt{2}}\phi_2(x) + \frac{1}{2}\phi_0(x). $$
We use these results to write each term in equation (34) as a sum of three-particle eigenstates with the appropriate coefficients. Collecting terms involving the same eigenstates, the final expression for $\Psi(0+)$ up to second order in $\alpha$ is then
\begin{equation}
\begin{aligned}
\label{expanded perturbative wave function immediately after interaction in terms of energy eigenstates}
\Psi(0+) = \frac{1}{\sqrt{2}} & \biggl( \left(1 - \frac{3\alpha^2}{8} \right)\phi_{010} -\frac{\alpha}{2}\phi_{001} + \frac{\alpha}{\sqrt{2}}\phi_{021} + \frac{\sqrt{3} \alpha^2}{4\sqrt{2}}\phi_{032} - \frac{3 \alpha^2}{4\sqrt{2}}\phi_{012} + \frac{\sqrt{3} \alpha^2}{8}\phi_{030} \\
& + i\left( \left(1 - \frac{\alpha^2}{8}\right)\phi_{100} + \frac{\alpha}{2}\phi_{111} + \frac{\alpha^2}{4}\phi_{122} - \frac{\alpha^2}{4 \sqrt{2}}\phi_{102} + \frac{\alpha^2}{4\sqrt{2}}\phi_{120} \right) \biggr).
\end{aligned}
\end{equation}

It can be seen that the norm of the above expression is equal to one, with corrections of order $\alpha^4$.

Now that we have an expression for $\Psi(0+)$ in terms of the eigenstates of the Hamiltonian, we can find the approximate wave function at any time $t$ by simply attaching the appropriate phase factors $e^{-iE_{jkl}t}$ to each eigenstate. Also removing overall phase factors, we then have
\begin{equation}
\begin{aligned}
\Psi(t) = \frac{1}{\sqrt{2}} & \biggl( \left(1 - \frac{3\alpha^2}{8} \right)e^{-it}\phi_{010} -\frac{\alpha}{2}e^{-it}\phi_{001} + \frac{\alpha}{\sqrt{2}}e^{-3it}\phi_{021} + \frac{\sqrt{3} \alpha^2}{4\sqrt{2}}e^{-5it}\phi_{032} \\
& - \frac{3 \alpha^2}{4\sqrt{2}}e^{-3it}\phi_{012} + \frac{\sqrt{3} \alpha^2}{8}e^{-3it}\phi_{030} \\
& + i\left( \left(1 - \frac{\alpha^2}{8}\right)e^{-it}\phi_{100} + \frac{\alpha}{2}e^{-3it}\phi_{111} + \frac{\alpha^2}{4}e^{-5it}\phi_{122} - \frac{\alpha^2}{4 \sqrt{2}}e^{-3it}\phi_{102} + \frac{\alpha^2}{4\sqrt{2}}e^{-3it}\phi_{120} \right) \biggr).
\end{aligned}    
\end{equation}

The velocity field is proportional to the gradient of the overall phase of the wave function. We use this to simulate the evolution of the system and obtain snapshots of the marginal distribution for A, as well as its equilibrium counterpart, at various final times. The simulation follows the backtracking method described in ref. \cite{VW05} to find the values of $\rho(t)$ and $\rho^{\textrm{eq}}(t) = |\Psi(t)|^2$ on a fine grid of uniformly-spaced points in the main support of the two distributions. The joint distribution $\rho$ is then integrated over the positions of particles B and C to obtain the marginal distribution $\rho_A$ for particle A. The equilibrium marginal distribution $\rho_A^{\textrm{eq}}$ remains constant throughout, always equal to one half of the sum of the distributions corresponding to the ground and first excited states. This last point follows entirely from the Schrödinger equation and is not affected by the presence of quantum nonequilibrium. However, since A is entangled with B and C following the interaction, the evolution of the actual marginal distribution $\rho_A$ depends on the velocity field in the three-particle configuration space. Since B and C are now out of equilibrium, this dependence causes particle A to evolve away from equilibrium.

\subsection{Numerical results}
 
As an illustrative example we choose $\alpha = 0.1$, so that the calculations are accurate to order $\alpha^2 = 0.01$ and the error is of order $\alpha^3 = 0.001$. The resulting marginal distributions for A, as well as the corresponding equilibrium marginal distributions, are shown for various times in Figure 5.

\begin{figure}[H]
    \begin{subfigure}{0.5\textwidth}
        \includegraphics[width = 6cm, height = 4.5cm]{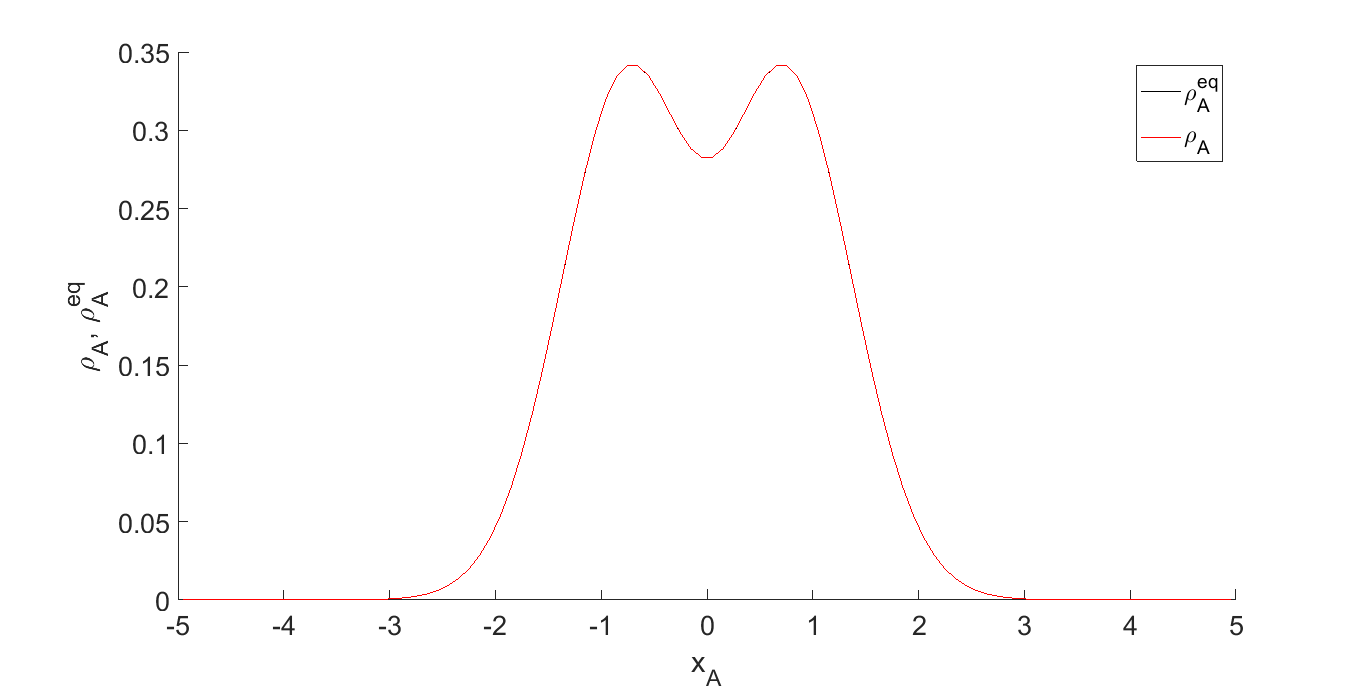}
        \subcaption{$t = 0+$}
    \end{subfigure}
    \begin{subfigure}{0.5\textwidth}
        \includegraphics[width = 6cm, height = 4.5cm]{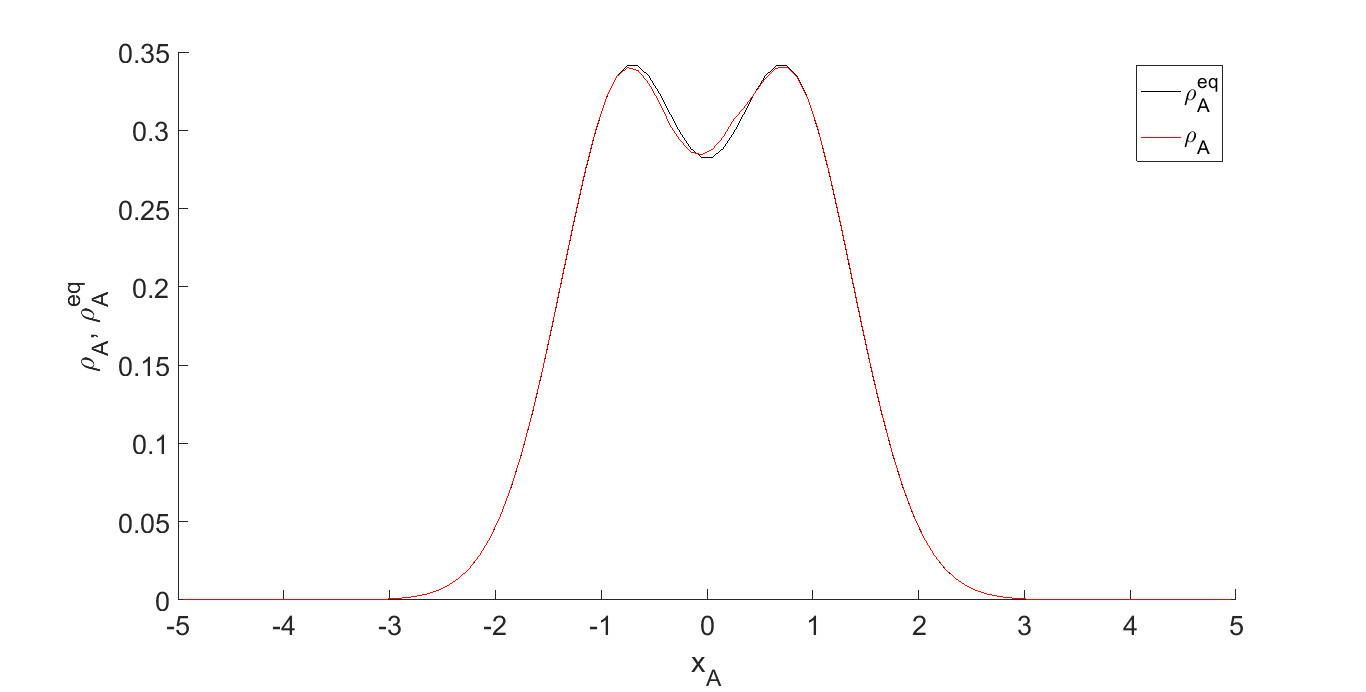}
        \subcaption{$t = T/16 $}
    \end{subfigure}
    \begin{subfigure}{0.5\textwidth}
        \includegraphics[width = 6cm, height = 4.5cm]{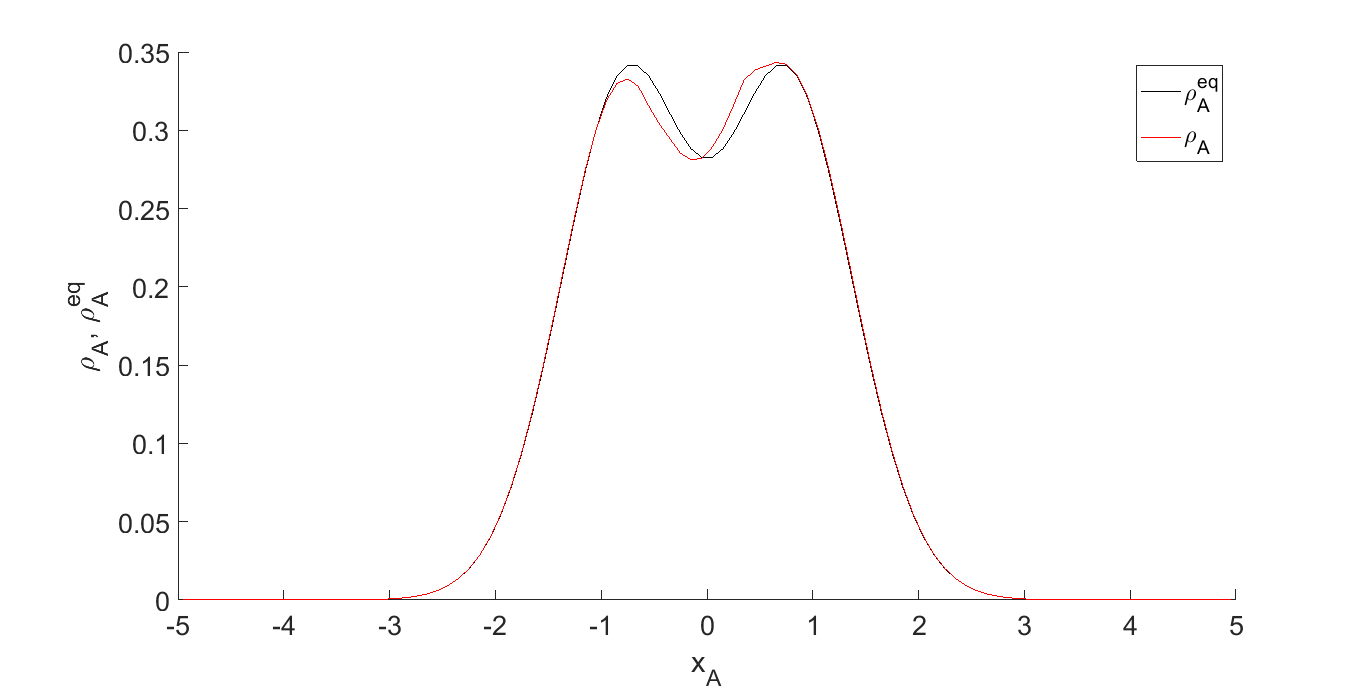}
        \subcaption{$t = T/8 $}
    \end{subfigure}
    \begin{subfigure}{0.5\textwidth}
        \includegraphics[width = 6cm, height = 4.5cm]{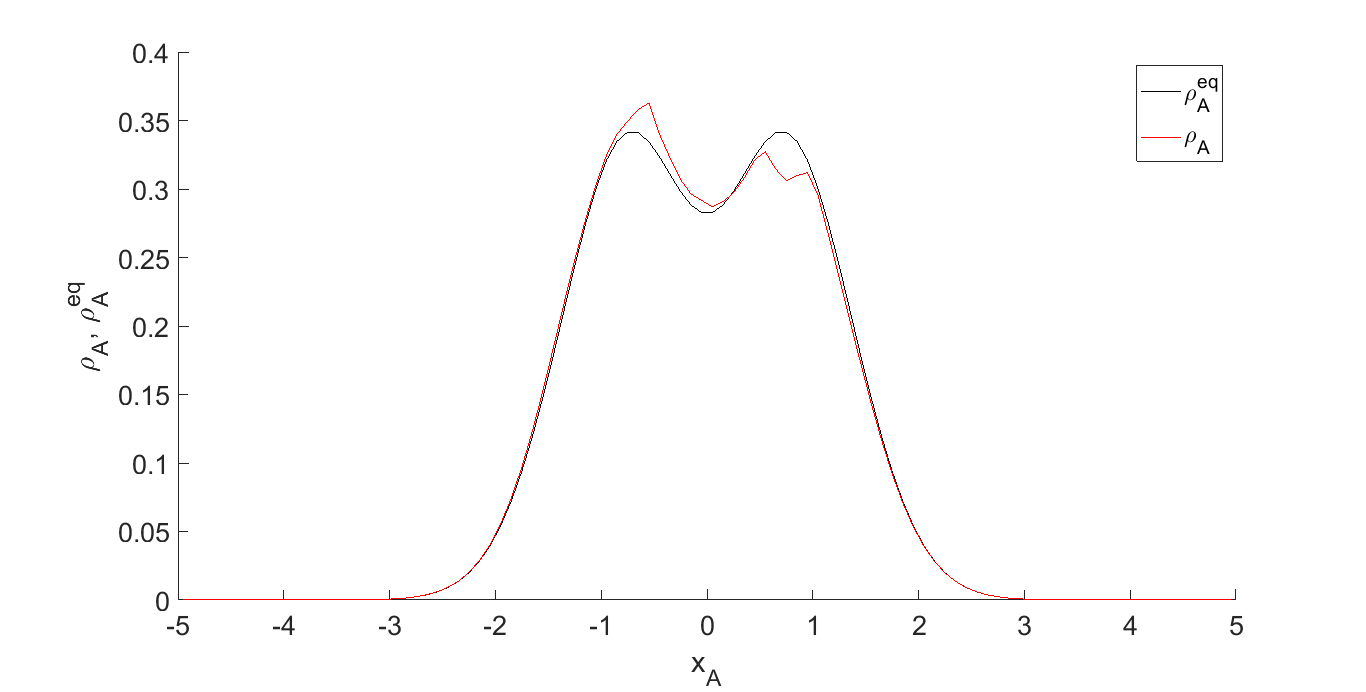}
        \subcaption{$t = T/2 $}
    \end{subfigure}
     \caption{Snapshots of the marginal and equilibrium marginal distributions for particle A, with the former in red and the latter in black, at the times $0+$, $T/16$, $T/8$ and $T/2$. Here $T = 2\pi$ refers to one period of each oscillators.}
\end{figure}

As can be seen from Figure 5, particle A is in equilibrium immediately after the interaction but is subsequently pulled out of equilibrium. The difference between the actual and equilibrium marginals for A peaks at about $0.03$ in Figure 5(d), which is about $10\%$ of the value of the distributions and clearly of higher order than the expected error $\alpha^3 = 0.001$. By contrast, we find that the normalization of the distributions is indeed accurate to within the expected error.

Thus we have demonstrated, through both analytical calculation and numerical simulation, that a quantum particle can be pulled out of equilibrium if its entangled partner interacts with another system that is already out of equilibrium.

\subsection{Extension to field modes}

In the above simulation we considered three harmonic oscillators with masses and angular frequencies set to unity. The results can be easily extended to (identical) oscillators with unit mass and general angular frequency $\omega$. We keep the mass set to unity because we are interested in the analogy with a scalar field mode, which is equivalent to an oscillator of unit mass and angular frequency $\omega = k$ (as we saw in Section 2.3).

For an oscillator with $m = 1$ and angular frequency $\omega$, the $n^{th}$ energy eigenstate $\phi_n(x)$ is given by
\begin{equation}
\label{energy eigenstate}
\phi_n(x) = \frac{1}{\sqrt{2^n n!}} \left(\frac{\omega}{\pi}\right)^{\frac{1}{4}} e^{-\frac{\omega x^2}{2}} H_n(\sqrt{\omega} x).
\end{equation}
Defining $u = \sqrt{\omega} x$, we can rewrite this as
\begin{equation}
\label{energy eigenstate in terms of u}
\phi_n(u) = \frac{1}{\sqrt{2^n n!}} \left(\frac{\omega}{\pi}\right)^{\frac{1}{4}} e^{-\frac{u^2}{2}} H_n(u).
\end{equation}
This can be compared with $\phi_n(x)$ with $\omega = 1$, which is identical to $\phi_n(u)$ with $u$ replaced by $x$ except for the missing normalization factor $\omega^{\frac{1}{4}}$.

Since we wish to extend our results to general oscillators, our simulation can be thought of as occurring on a lattice of $u_A$, $u_B$ and $u_C$. The wave function immediately after the interaction will be as in eqnuation (\ref{expanded perturbative wave function immediately after interaction in terms of energy eigenstates}), with the understanding that the $\phi_{jkl}$'s are the same functions as before with $x_A$, $x_B$ and $x_C$ replaced by $u_A$, $u_B$ and $u_C$. This wave function is not normalized with respect to $x_A$, $x_B$ and $x_C$ because it is missing a factor $\omega^{\frac{1}{4}}$ for each particle. However, this does not affect the phase $S$ of the wave function, which is what determines the velocity field.

In our above simulation, each term in the wave function has a phase factor $e^{iE_{jkl}t}$ at time $t$, with $E_{jkl} = j + k + l + \frac{3}{2}$ being the energies of three oscillators of angular frequency 1. However, when the angular frequency is $\omega$, the energies are actually $E^\prime_{jkl} = \omega \left(j + k + l + \frac{3}{2}\right)$. Hence the phase factor in each term of the wave function is really $\omega$ times the value of the previous corresponding phase factor. To compensate for this we may work with the rescaled time variable $t^\prime = \omega t$, so that the phase factors are indeed $e^{iE_{jkl}t^\prime}$ with $E_{jkl} = j + k + l + \frac{3}{2}$ as before. As it turns out, this transformation also allows us to use the same guidance equation as before.

In terms of $u$ the velocity field is given by

$$ \frac{d u}{d t} = \sqrt{\omega} \frac{d x}{d t} = \sqrt{\omega} \frac{\partial S}{\partial x} = \omega \frac{\partial S}{\partial u}. $$ 
Our simulation is set up to solve the differential equation $\frac{d u}{d t} = \frac{\partial S}{\partial u}$. However, if we replace $t$ with $ t^\prime = \omega t$, the guidance equation for $u$ in terms of $t^\prime$ is given by 
$$\frac{d u}{d t^\prime} = \frac{\partial S}{\partial u}.$$ Thus, our simulation calculates the wave function at values of $t^\prime$ and evolves the trajectories according to the velocity field in terms of $t^\prime$.

Now we need to find the analogous nonequilibrium distribution of particle C. In our simulation, the nonequilibrium distribution now reads

$$ \rho_C(u_C, 0) = \frac{1}{\sqrt{\pi}} e^{-(u_C - 2)^2}. $$
This is the same as 
$$\rho_C (x_C, 0) = \frac{1}{\sqrt{\pi}} e^{-\omega (x_C - \frac{2}{\sqrt{\omega}})^2},$$ 
which means that the corresponding nonequilibrium distribution for particle C is a Gaussian with the same width as the ground state but with mean shifted by a distance $d = \frac{2}{\sqrt{\omega}}$. The initial distributions for particles A and B are of course their equilibrium distributions, given by the squared amplitude of their initial wave functions. The overall initial distribution is again missing a factor of $\omega^{\frac{1}{2}}$ for each particle and hence is not normalized, but this is as it should be (if the wave function is missing factors of $\omega^{\frac{1}{4}}$, the Born distribution corresponding to the wave function is missing factors of $\omega^{\frac{1}{2}}$). 

Thus, to conclude, a snapshot of our simulation at time $t^\prime$ corresponds to real time $t = \frac{t^\prime}{\omega}$. The results in Figure 5 can still be interpreted as indicating the difference between the marginal distributions of particle A at various final times, with the understanding that $T = \frac{2 \pi}{\omega}$ is the period of the oscillator. For example, Figure 5(d) is a snapshot at $t^\prime = \pi$, which means the real time for the oscillator corresponding to this snapshot is $t = \frac{\pi}{\omega} = \frac{T}{2}$.

Hence, when $\alpha = 0.1$, the percentage difference between the values of the marginal distributions is about $10\%$ near the peaks of the equilibrium marginal at $t = \frac{T}{2} = \frac{\pi}{\omega}$. This applies for the analogous problem of a field mode degree of freedom with $\omega = k$ and the initial conditions described above. 

\section{Conclusion}

We have shown by example, both analytically and numerically, that a system initially in quantum equilibrium can evolve away from equilibrium if it is entangled with another equilibrium system that is interacting locally with a third system that is initially out of equilibrium. In other words, quantum nonequilibrium can propagate nonlocally across the light cone.

While our calculations strictly apply only to decoupled field modes on Minkowski spacetime, the formalism of Section 2.2 shows how similar calculations can in principle be carried out on a background curved spacetime, yielding the result that quantum nonequilibrium can propagate nonlocally across an event horizon. While this effect is of interest in its own right, our motivation for studying it comes from the proposal \cite{AVarx2, AV07} that it could provide a mechanism whereby information can escape from the interior of a black hole to the exterior, despite the usual limits associated with standard causal horizons. As noted, this scenario requires there to be interior degrees of freedom that are already out of quantum equilibrium -- presumably owing to Planck-scale physics operating near the singularity \cite{AV10, AVPIRSA, AVprep}.

The results reported here provide a proof-of-concept for the scenario first proposed in refs. \cite{AVarx2, AV07}. The mechanism illustrated here provides a means whereby information can be transmitted from the interior of a black hole to the exterior region. In principle, then, the mechanism provides an alternative approach whereby we might attempt to avoid information loss in an evaporating black hole. In practice, however, many questions remain unanswered.

Firstly, it remains to calculate how much information could reasonably escape to the exterior during the lifetime of an evaporating black hole. In the examples studied here, for ease of calculation we have resorted to perturbation theory, so that the effects are necessarily small. But there seems to be no reason to expect them to remain small outside of the perturbative regime. How large the effects can realistically be remains to be studied.

Secondly, the size of the effects will also depend on the magnitude of the nonequilibrium that exists behind the black-hole horizon. To estimate this will require a detailed model of the interior, and in particular of the expected Planck-scale effects close to the singularity. Recent work in quantum gravity provides a model of the gravitational instability of the Born rule during cosmological inflation \cite{AVPIRSA, AVprep}. How to extend such models to black-hole singularities is however a subject for future work.


\begin{thebibliography}{99}

\bibitem{deB28} L. de Broglie, La nouvelle dynamique des quanta, in: \textit{\'{E}lectrons et
Photons: Rapports et Discussions du Cinqui\`{e}me Conseil de Physique
}(Gauthier-Villars, Paris, 1928). [English translation in ref. \cite{BV09}.]

\bibitem{BV09} G. Bacciagaluppi and A. Valentini, \textit{Quantum Theory at the Crossroads:
Reconsidering the 1927 Solvay Conference} (Cambridge University Press, 2009). [arXiv:quant-ph/0609184]

\bibitem{B52a} D. Bohm, A suggested interpretation of the quantum theory in terms of `hidden'
variables. I, Phys. Rev. \textbf{85}, 166 (1952).

\bibitem{B52b} D. Bohm, A suggested interpretation of the quantum theory in terms of `hidden'
variables. II, Phys. Rev. \textbf{85}, 180 (1952).

\bibitem{Holl93} P. R. Holland, \textit{The Quantum Theory of Motion: an Account of the de
Broglie-Bohm Causal Interpretation of Quantum Mechanics} (Cambridge University
Press, Cambridge, 1993).

\bibitem{AV07} A. Valentini, Astrophysical and cosmological tests of quantum theory, J. Phys.
A: Math. Theor. \textbf{40}, 3285 (2007). [arXiv:hep-th/0610032]

\bibitem{AV91a} A. Valentini, Signal-locality, uncertainty, and the subquantum \textit{H}%
-theorem. I, Phys. Lett. A \textbf{156}, 5 (1991).

\bibitem{AV91b} A. Valentini, Signal-locality, uncertainty, and the subquantum \textit{H}%
-theorem. II, Phys. Lett. A \textbf{158}, 1 (1991).

\bibitem{AV92} A. Valentini, On the pilot-wave theory of classical, quantum and subquantum
physics, PhD thesis, International School for Advanced Studies, Trieste, Italy (1992). [http://hdl.handle.net/20.500.11767/4334]

\bibitem{AV01} A. Valentini, Hidden variables, statistical mechanics and the early universe,
in: \textit{Chance in Physics: Foundations and Perspectives}, eds. J. Bricmont
\textit{et al}. (Springer, Berlin, 2001). [arXiv:quant-ph/0104067]

\bibitem{VW05} A. Valentini and H. Westman, Dynamical origin of quantum probabilities, Proc.
Roy. Soc. Lond. A \textbf{461}, 253 (2005). [arXiv:quant-ph/0403034]

\bibitem{TRV12} M. D. Towler, N. J. Russell, and A. Valentini, Time scales for dynamical
relaxation to the Born rule, Proc. Roy. Soc. Lond. A \textbf{468}, 990 (2012). [arXiv:1103.1589]

\bibitem{SC12} S. Colin, Relaxation to quantum equilibrium for Dirac fermions in the de
Broglie--Bohm pilot-wave theory, Proc. Roy. Soc. Lond. A \textbf{468}, 1116
(2012). [arXiv:1108.5496]

\bibitem{ACV14} E. Abraham, S. Colin and A. Valentini, Long-time relaxation in pilot-wave
theory, J. Phys. A: Math. Theor. \textbf{47}, 395306 (2014). [arXiv:1310.1899]

\bibitem{AV02} A. Valentini, Subquantum information and computation, Pramana -- J. Phys.
\textbf{59}, 269 (2002). [arXiv:quant-ph/0203049]

\bibitem{AV04} A. Valentini, Universal signature of non-quantum systems, Phys. Lett. A
\textbf{332}, 187 (2004). [arXiv:quant-ph/0309107]

\bibitem{AV09} A. Valentini, Beyond the quantum, Physics World \textbf{22N11}, 32 (2009). [arXiv:1001.2758]

\bibitem{PV06} P. Pearle and A. Valentini, Quantum mechanics: generalizations, in:
\textit{Encyclopaedia of Mathematical Physics}, eds. J.-P. Fran\c{c}oise
\textit{et al}. (Elsevier, North-Holland, 2006). [arXiv:quant-ph/0506115]

\bibitem{AV96} A. Valentini, Pilot-wave theory of fields, gravitation and cosmology, in:
\textit{Bohmian Mechanics and Quantum Theory: an Appraisal}, eds. J. T.
Cushing \textit{et al}. (Kluwer, Dordrecht, 1996).

\bibitem{AV10} A. Valentini, Inflationary cosmology as a probe of primordial quantum
mechanics, Phys. Rev. D \textbf{82}, 063513 (2010). [arXiv:0805.0163]

\bibitem{UV15} N. G. Underwood and A. Valentini, Quantum field theory of relic nonequilibrium
systems, Phys. Rev. D \textbf{92}, 063531 (2015). [arXiv:1409.6817]

\bibitem{DGZ92} D. D\"{u}rr, S. Goldstein and N. Zangh\`{\i}, Quantum equilibrium and the
origin of absolute uncertainty, J. Stat. Phys. \textbf{67}, 843 (1992). [arXiv:quant-ph/0308039]

\bibitem{Tum18} R. Tumulka, Bohmian mechanics, in: \textit{The Routledge Companion to the
Philosophy of Physics}, eds. E. Knox and A. Wilson (Routledge, 2018). [arXiv:1704.08017]

\bibitem{Allori} A. Valentini, Foundations of statistical mechanics and the status of the Born
rule in de Broglie-Bohm pilot-wave theory, ArXiv:1906.10761. To appear in: \textit{Statistical Mechanics and Scientific Explanation: Determinism, Indeterminism and Laws of Nature}, ed. V. Allori (World Scientific).

\bibitem{AVarx1} A. Valentini, De Broglie-Bohm prediction of quantum violations for
cosmological super-Hubble modes, arXiv:0804.4656.

\bibitem{CV13} S. Colin and A. Valentini, Mechanism for the suppression of quantum noise at
large scales on expanding space, Phys. Rev. D \textbf{88}, 103515 (2013). [arXiv:1306.1579]

\bibitem{CV15} S. Colin and A. Valentini, Primordial quantum nonequilibrium and large-scale
cosmic anomalies, Phys. Rev. D \textbf{92}, 043520 (2015). [arXiv:1407.8262]

\bibitem{CV16} S. Colin and A. Valentini, Robust predictions for the large-scale cosmological
power deficit from primordial quantum nonequilibrium, Int. J. Mod. Phys. D
\textbf{25}, 1650068 (2016). [arXiv:1510.03508]

\bibitem{SPV19} S. Vitenti, P. Peter and A. Valentini, Modeling the large-scale power deficit
with smooth and discontinuous primordial spectra, Phys. Rev. D \textbf{100},
043506 (2019). [arXiv:1901.08885]

\bibitem{AVarx2} A. Valentini, Black holes, information loss, and hidden variables, arXiv:hep-th/0407032.

\bibitem{Hawk76} S. W. Hawking, Breakdown of predictability in gravitational collapse, Phys.
Rev. D \textbf{14}, 2460 (1976).

\bibitem{Polch17} J. Polchinski, The black hole information problem, in: \textit{New Frontiers
in Fields and Strings} (World Scientific, 2017). [arXiv:1609.04036]

\bibitem{BD82} N. D. Birrell and P. C. W. Davies, \textit{Quantum Fields in Curved Space}
(Cambridge University Press, Cambridge, 1982).

\bibitem{Wald94} R. M. Wald, \textit{Quantum Field Theory in Curved Spacetime and Black Hole
Thermodynamics} (University of Chicago Press, Chicago, 1994).

\bibitem{Gidd06} S. B. Giddings, Black hole information, unitarity, and nonlocality, Phys. Rev.
D \textbf{74}, 106005 (2006). [arXiv:hep-th/0605196]

\bibitem{AVPIRSA} A. Valentini, Is the Born rule unstable in quantum gravity?, seminar at
Perimeter Institute (2018), http://pirsa.org/18020103.

\bibitem{AVprep} A. Valentini, Quantum-gravitational instability of the Born rule, arXiv, to appear.

\bibitem{CKS16} B. Carr, F. Kuhnel and M. Sandstad, Primordial black holes as dark matter,
Phys. Rev. D \textbf{94}, 083504 (2016). [arXiv:1607.06077]

\bibitem{ADM} R. Arnowitt, S. Deser and C. W. Misner, The dynamics of general relativity,
in: \textit{Gravitation: an Introduction to Current Research}, ed. L. Witten
(Wiley, 1962).

\bibitem{HE73} S. W. Hawking and G. F. R. Ellis, \textit{The Large Scale Structure of
Space-Time} (Cambridge University Press, Cambridge, 1973).

\bibitem{AV08a} A. Valentini, Hidden variables and the large-scale structure of space-time,
in: \textit{Einstein, Relativity and Absolute Simultaneity}, eds. W. L. Craig
and Q. Smith (Routledge, London, 2008). [arXiv:quant-ph/0504011]

\bibitem{SV09} W. Struyve and A. Valentini, De Broglie-Bohm guidance equations for arbitrary Hamiltonians, J. Phys. A: Math. Theor. \textbf{42}, 035301 (2009). [arXiv:0808.0290]

\end{thebibliography}
\end{document}